\documentclass{PoS}
\usepackage{epsfig,amsmath,amssymb}

\pdfoutput=1 

\newcommand{\beq}{\begin{equation}}
\newcommand{\eeq}{\end{equation}}
\newcommand{\bea}{\begin{eqnarray}}
\newcommand{\eea}{\end{eqnarray}}
\newcommand{\Fig}[1]{Fig.\,\ref{#1}}

\newcommand{\Eq}[1]{Eq.\,(\ref{#1})}

\newcommand{\Eqsand}[2]{Eqs.\,(\ref{#1}) and (\ref{#2})}

\newcommand{\f}{\frac}

\newcommand{\as}{\alpha_s}
\newcommand{\aem}{\alpha}

\newcommand{\MW}{M_{\scriptscriptstyle W}}
\newcommand{\MZ}{M_{\scriptscriptstyle Z}}

\newcommand{\mc}{m_c}
\newcommand{\mb}{m_b}
\newcommand{\mt}{m_t}

\newcommand{\muc}{\mu_c}

\newcommand{\mut}{\mu_t}

\newcommand{\GF}{G_F}
\newcommand{\BR}{{\rm BR}}
\newcommand{\MeV}{{\rm MeV}}
\newcommand{\GeV}{{\rm GeV}}

\newcommand{\MSbar}{\overline{\rm MS}}
\newcommand{\re}{{\rm Re}}
\newcommand{\im}{{\rm Im}}
\newcommand{\ord}{O}
\def\unit{\leavevmode\hbox{\small1\kern-3.6pt\normalsize1}}
\newcommand{\eps}{\epsilon}

\newcommand{\sL}{{\scalebox{0.6}{$L$}}}
\newcommand{\sR}{{\scalebox{0.6}{$R$}}}
\newcommand{\Kpnn}{K^+ \to \pi^+ \nu \bar{\nu}}
\newcommand{\Kpnng}{K^+ \to \pi^+ \nu \bar{\nu} (\gamma)}
\newcommand{\KLnn}{K_L \to \pi^0 \nu \bar{\nu}}
\newcommand{\Knns}{K \to \pi \nu \bar{\nu}}

\newcommand{\Kpien}{K \to \pi \ell \bar{\nu}}
\newcommand{\Klthree}{K_{\ell 3}}
\newcommand{\bXpartg}{b \to X^{\rm partonic}_s \gamma}

\newcommand{\BXsga}{\bar{B} \to X_s \gamma}

\newcommand{\BXsll}{\bar{B} \to X_s \ell^+ \ell^-}

\newcommand{\BRKpg}{\BR (\Kpnn (\gamma))}
\newcommand{\BRKL}{\BR (\KLnn)}

\newcommand{\BRga}{\BR (\BXsga)}
\newcommand{\BRll}{\BR (\BXsll)}

\newcommand{\stodvv}{s \to d \nu \bar{\nu}}
\newcommand{\btosg}{b \to s \gamma}

\newcommand{\btosll}{b \to s \ell^+ \ell^-}
\newcommand{\Ztobb}{Z \to b \bar{b}}

\newcommand{\Rb}{R_b^0}
\newcommand{\Ab}{A_b}
\newcommand{\AFB}{A_{\rm FB}^{0, b}}

\newcommand{\CP}{C\hspace{-0.25mm}P}

\newcommand{\mysigma}{\hspace{0.4mm} \sigma}
\newcommand{\Pc}{P_c}
\newcommand{\dPcu}{\delta P_{c,u}}

\newcommand{\etal}{{\it et al}.}

\title{\boldmath Rare \scalebox{1.1}{$K$}- ({\it vs}.)
  \scalebox{1.1}{$B$}-decays \unboldmath}

\ShortTitle{\boldmath Rare $K$- ({\it vs}.) $B$-decays \unboldmath}

\author{\speaker{Ulrich Haisch}\thanks{We thank the organizers for the
    pleasant and stimulating atmosphere they were able to create
    during ``KAON'07''. Private communications with Einan Gardi and
    Miko{\l}aj Misiak concerning $\BXsga$ are acknowledged. This work
    is supported by the Schweizer Nationalfonds.} \\
  Institut f\"ur Theoretische Physik, Universit\"at Z\"urich,
  CH-8057 Z\"urich, Switzerland\\
  E-mail: \email{uhaisch@physik.unizh.ch}}

\abstract{We present a concise review of the recent theoretical
  progress concerning the standard model calculations of the rare
  $\KLnn$, $\Kpnng$, $\BXsga$, and $\BXsll$ decays. The current status
  and future of the model-independent analysis of rare $K$- and
  $B$-meson decays within constrained minimal-flavor-violation is also
  briefly discussed.}

\FullConference{KAON International Conference\\
		May 21-25 2007\\
		Laboratori Nazionali di Frascati dell'INFN, Rome, Italy}

\begin{document}

\section{\boldmath Warm-up: basic facts about $\stodvv$ and $\btosg$
  \unboldmath}

\begin{figure}[!t]
\begin{center}
\includegraphics[height=2.5cm]{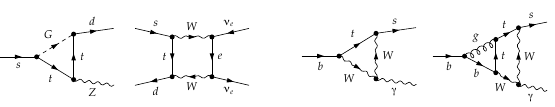} 
\begin{picture}(0,0)(0,0)
\put(-347,-10){$\propto \f{\mt^2}{\MW^2}$}
\put(-262,-10){$\propto {\rm const.}$}
\put(-150,-10){$\propto {\rm const.}$}
\put(-65,-10){$\propto \as \ln \f{\mb}{\MW}$}
\end{picture}
\vspace{4mm}
\caption{Feynman diagrams contributing to the $\stodvv$ (left) and
  $\btosg$ (right) transition in the SM. Their scaling behavior in the
  heavy top quark mass limit is also shown. See text for details.}
\label{fig:penguins&boxes}
\end{center}
\end{figure}

The $\stodvv$ transition is one of the rare examples of an electroweak
(EW) process whose leading contribution starts at $\ord (\GF^2)$
within the standard model (SM). At the one-loop level it proceeds
through $Z$-penguin and EW box diagrams which are highly sensitive to
the underlying short-distance (SD) dynamics. Sample diagrams are shown on
the left of \Fig{fig:penguins&boxes}. Separating the contributions
according to the intermediate up-type quark running inside the loops,
the QCD corrected amplitude takes the form
\beq \label{eq:Asdvv}
A_{\rm SM} (\stodvv) = \sum_{q = u, c, t} V_{qs}^\ast V_{qd} X_{\rm
  SM}^q \propto \f{\mt^2}{\MW^2} (\lambda^5 + i \lambda^5) +
\f{\mc^2}{\MW^2} \ln \f{\mc}{\MW} \lambda + \f{\Lambda^2}{\MW^2}
\lambda \, ,
\eeq
where $V_{ij}$ denote the elements of the Cabibbo-Kobayashi-Maskawa
(CKM) matrix and $\lambda = |V_{us}| = 0.225$. The hierarchy of the
CKM elements would obviously favor the charm and up quark
contributions, but the power-like Glashow-Iliopoulos-Maiani (GIM)
mechanism, arising mainly from the $SU(2)_L$ breaking in the
$Z$-penguin amplitude, leads to a very different picture. The top
quark contribution, carrying a large $\CP$-violating phase, accounts
for $\sim 68 \%$ of $A_{\rm SM} (\stodvv)$, while corrections due to
internal charm and up quarks amount to $\sim 29 \%$ and a mere $\sim 3
\%$. These properties imply that long-distance (LD) effects in the
direct $\CP$-violating $\KLnn$ mode are negligible, while they are
highly suppressed in $\Kpnng$.

A related important feature, following from the EW structure of
$A_{\rm SM} (\stodvv)$ as well, is that the SD contributions to both
$\Knns$ decay modes are governed by a single effective operator,
namely
\beq \label{eq:Qv}
Q_\nu = (\bar s_{\sL} \gamma_\mu d_{\sL}) (\bar \nu_{\sL} \gamma^\mu
\nu_{\sL}) \, . 
\eeq
By virtue of the conservation of the $V-A$ current, large QCD
logarithms appear only in the charm quark contribution to
\Eq{eq:Asdvv}. The hadronic matrix element of $Q_\nu$ itself, can be
extracted very precisely from the wealth of available data on $\Kpien$
($\Klthree$) decays.

In summary, the superb theoretical cleanness and the enhanced
sensitivity to both non-standard flavor and $\CP$ violation, make the
$\stodvv$ channels unique tools to discover or, if no deviation is
found, to set severe constraints on non-minimal-flavor-violating (MFV)
physics where the hard GIM cancellation present in the SM and MFV is
in general no longer active \cite{Cecilia&Chris}. 

Unlike $\stodvv$, the $\btosg$ transition is dominated by perturbative
QCD effects which replace the power-like GIM mechanism present in the
EW vertex by a logarithmic one. The mild suppression of the QCD
corrected SM amplitude
\beq \label{eq:Absg}
A_{\rm SM} (\btosg) = \sum_{q = u, c, t} V_{qb}^\ast V_{qs} K_{\rm
  SM}^q \propto \ln \f{\mb}{\MW} \lambda^2 + \ln \f{\mb}{\MW}
\lambda^2 + \ln \f{\mb}{\MW} \lambda^4 \, , 
\eeq
reduces the sensitivity of the process to high scale physics, but
enhances the $\BXsga$ branching ratio (BR) with respect to the purely
EW prediction by a factor of around three. The logarithmic GIM
cancellation originates from the non-conservation of the effective
tensor operator 
\beq \label{eq:Q7} 
Q_7 = \f{e}{16 \pi^2} \mb (\bar s_{\sL} \sigma_\mu b_{\sR}) F^{\mu
  \nu} \, ,
\eeq 
which is generated at the EW scale by photon penguin diagrams
involving $W$-boson and top quark exchange. Sample one- and two-loop
diagrams are shown on the right of \Fig{fig:penguins&boxes}.

After including logarithmic enhanced QCD effects, the dominant
contribution to the partonic $\bXpartg$ decay rate stems from charm
quark loops that amount to $\sim 158 \%$ of $A_{\rm SM} (\btosg)$. The
top contribution is with $\sim -60 \%$ of $A_{\rm SM} (\btosg)$
compared to the charm quark effects less than half as big and has the
opposite sign. Diagrams involving up quarks are suppressed by small
CKM factors and lead to an effect of only $\sim 2\%$ in $A_{\rm SM}
(\btosg)$. Due to the inclusive character of the $\BXsga$ mode and the
heaviness of the bottom quark, $\mb \gg \Lambda \sim \Lambda_{\rm
  QCD}$, non-perturbative effects arise only as small corrections to
the partonic decay rate.

In summary, the good theoretical control on and experimental
accessibility of the $\btosg$ rate, and its large generic sensitivity
to non-standard sources of flavor and $\CP$ violation, allows to
derive stringent constraints on a variety of new physics (NP) models,
in particular on those where the flavor-violating chiral transition
of the amplitude is not suppressed.

\section{\boldmath Recent theoretical progress in $\KLnn$ and $\Kpnng$ \unboldmath}

After summation over the three lepton families the SM BRs of the
$\Knns$ modes read 
\bea 
\label{eq:BRKL} \BRKL_{\rm SM} = \kappa_L \left ( \f{\im \,
    \lambda_t}{\lambda^5} \, X \right )^2 \, , \hspace{4cm}
\\ \label{eq:BRKp} \BRKpg_{\rm SM} = \kappa_+ (1 + \Delta_{\rm EM})
\left [ \left ( \f{\im \, \lambda_t}{\lambda^5} \, X \right )^2 +
  \left ( \f{\re \, \lambda_t}{\lambda^5} \, X + \f{\re
      \lambda_c}{\lambda} \left ( \Pc + \dPcu \right ) \right)^2
\right ] , \hspace{5mm}
\eea
where $\lambda_i = V_{is}^\ast V_{id}$. The top quark contribution $X
= 1.456 \pm 0.017_{\mt} \pm 0.013_{\mut} \pm 0.015_{\rm EW}$ is known
through next-to-leading order (NLO) in QCD \cite{X}. Its overall
uncertainty of $\sim 2 \%$ is in equal shares due to the parametric
error on the top quark mass, the sensitivity on the matching scale
$\mut$, and two-loop EW effects for which only the leading term in the
heavy top quark mass expansion has been calculated
\cite{Buchalla:1997kz}.

Major theoretical progress has been recently made concerning the
extraction of the hadronic $\langle \pi^i | \bar s \gamma_\mu d | K^j
\rangle$ matrix elements from $\Klthree$ data, by extending the
classic chiral perturbation theory (ChPT) analysis of leading order
(LO) $\ord (p^2 \eps^{(2)})$ isospin-breaking effects
\cite{Marciano:1996wy}. Here $\eps^{(2)} \propto (m_u - m_d)/m_s$. The
inclusion of NLO $\ord (p^4 \eps^{(2)})$ and partial
next-to-next-to-leading order (NNLO) $\ord (p^6 \eps^{(2)})$
corrections in the ChPT expansion \cite{Mescia:2007kn} leads to a
reduction of the uncertainties on the $\KLnn$ and $\Kpnn$ matrix
elements by a factor of $\sim 4$ and $\sim 7$. Since the overall
uncertainties on $\kappa_L = (2.229 \pm 0.017) \times 10^{-10} \,
(\lambda/0.225)^8$ and $\kappa_+ = (5.168 \pm 0.025) \times 10^{-10}
\, (\lambda/0.225)^8$ \cite{Mescia:2007kn} are now dominated by
experimental errors, a further improvement in the extraction of the
rare $K$-decay matrix elements will be possible with improved data for
the $\Klthree$ slopes and $K^+_{\ell 3}$ BRs. LD QED corrections
affecting $\Kpnng$ are encoded by $\Delta_{\rm EM}$ in
\Eq{eq:BRKp}. At LO in the ChPT expansion, these infrared finite $\ord
(p^2 \aem)$ corrections amount to $\Delta_{\rm EM} = -0.003$
\cite{Mescia:2007kn} for a maximum energy of $20 \, \MeV$ of the
undetected photon. More details on the ChPT analysis of rare $K$-decay
matrix elements can be found in \cite{Mescia:2007kn}.

\begin{figure}
\begin{center} 
\mbox{
\includegraphics[width=2.75in]{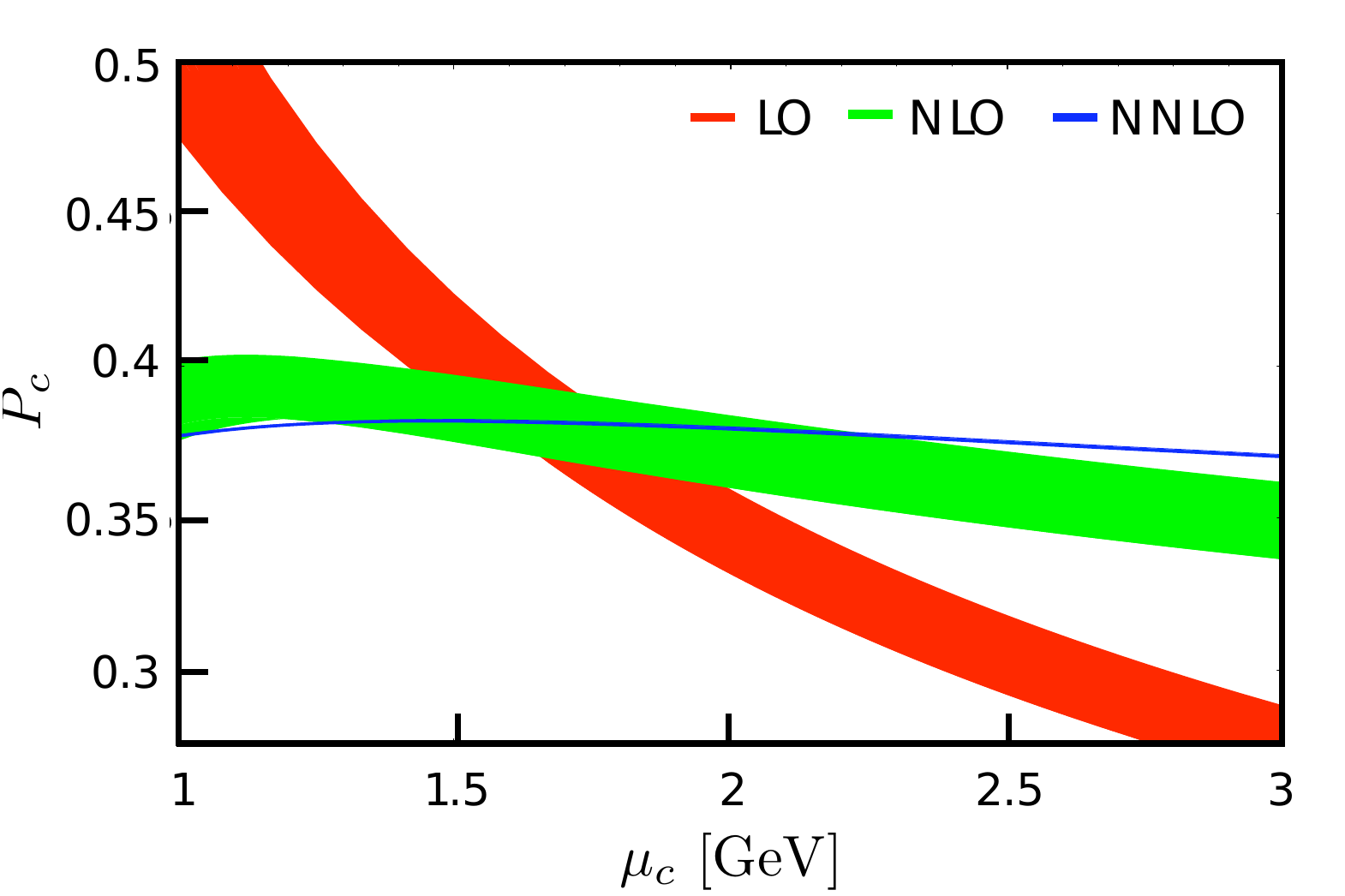}
}
\hspace{2.5mm}
\mbox{
\includegraphics[width=2.75in]{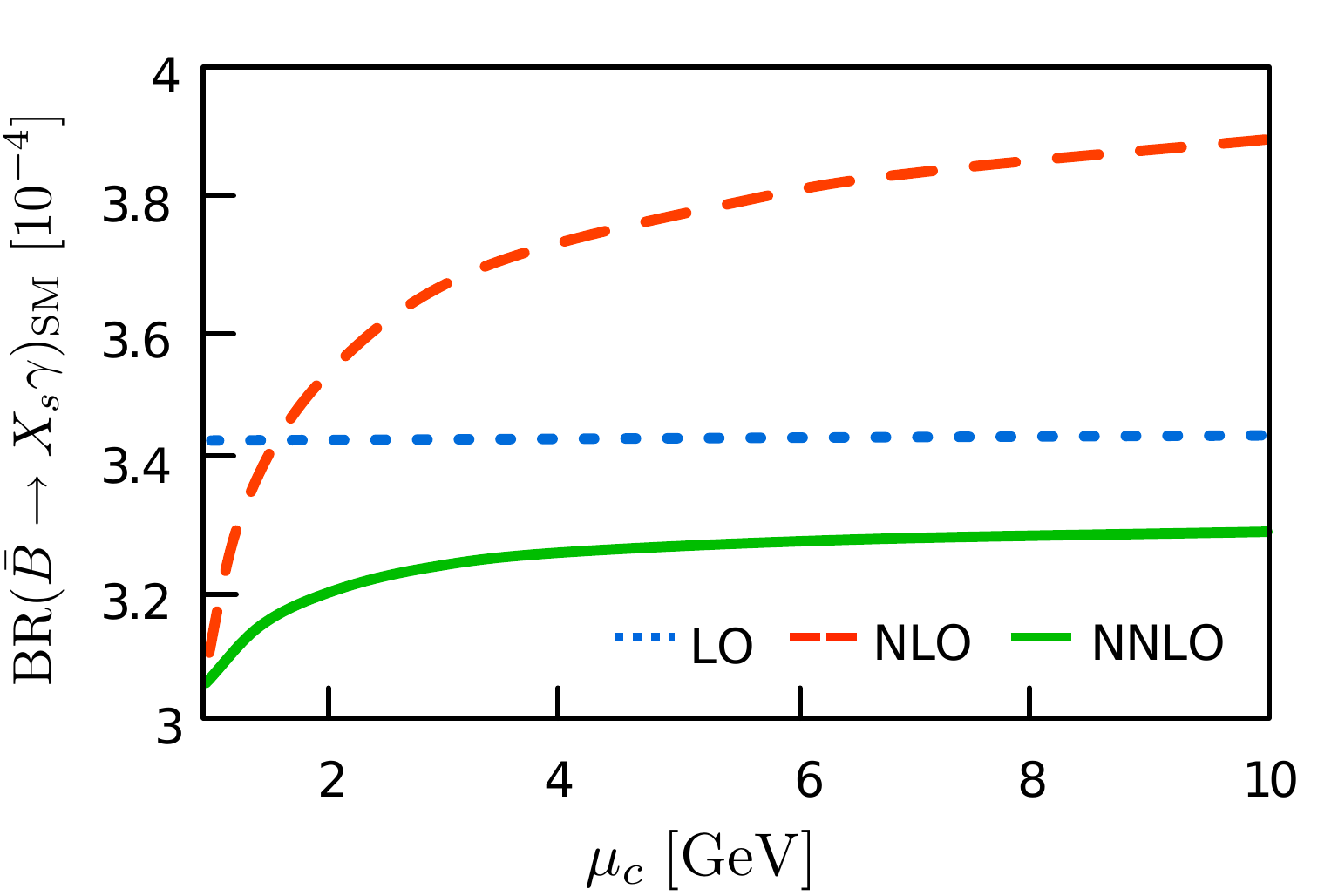}
}
\end{center}
\vspace{-6mm}
\caption{Charm quark mass renormalization scale $\muc$ dependence of
  $\Pc$ (left) and $\BRga_{\rm SM}$ (right) at LO, NLO, and NNLO in
  QCD. The width of the curves on the left indicate the uncertainty
  due to higher order terms in $\as$ that affect the evaluation of
  $\as (\muc)$ from $\as (\MZ)$. See text for details.}
\label{fig:scales}
\end{figure}

The parameter $\Pc$ entering \Eq{eq:BRKp} results from $Z$-penguin and
EW box diagrams involving internal charm quark exchange. As now both
high- and low-energy scales are involved, a complete renormalization
group analysis of this term is required. In this manner, large
logarithms $\ln \mc/\MW$ are resummed to all orders in $\as$. The
inclusion of NNLO QCD corrections \cite{NNLOPc} leads to a significant
reduction of the theoretical uncertainty by a factor of $\sim 4$, as
it removes almost the entire sensitivities of $\Pc$ on the charm quark
mass renormalization scale $\muc$ and on higher order terms in $\as$
that affect the evaluation of $\as (\muc)$ from $\as (\MZ)$. This is
illustrated by the plot in the left panel of \Fig{fig:scales}. For
$\mc = (1.30 \pm 0.05) \, \GeV$ one obtains the NNLO value $\Pc =
(0.374 \pm 0.031_{\mc} \pm 0.009_{\rm pert} \pm 0.009_{\as}) \,
(0.225/\lambda)^4$ \cite{NNLOPc}, where the individual errors are due
to the uncertainty on the charm quark $\MSbar$ mass, higher-order
perturbative effects, and the parametric error on $\as (\MZ)$. Since
the residual error on $\Pc$ is now fully dominated by the parametric
uncertainty coming from $\mc$, a better determination of the charm
quark mass is clearly an important theoretical goal in connection with
$\Kpnng$.

To gain an accuracy on the $\Kpnng$ BR of a few percent, it is
necessary to account for subleading effects not described by the
effective Hamiltonian that includes the dimension-six operator $Q_\nu$
of \Eq{eq:Qv}. The subleading corrections can be divided into two
groups: $i)$ contributions of dimension-eight four fermion operators
generated at the charm quark mass renormalization scale $\muc$
\cite{Falk:2000nm, Isidori:2005xm}, and $ii)$ genuine LD contributions
due to up quark loops which can be described within the framework of
ChPT \cite{Isidori:2005xm}. Both contributions can be effectively
included by $\dPcu = 0.04 \pm 0.02$ \cite{Isidori:2005xm} in
\Eq{eq:BRKp}. Numerically, they lead to an enhancement of $\BRKpg_{\rm
  SM}$ by $\sim 7 \%$. The quoted residual error of $\dPcu$ could in
principle be reduced by means of a dedicated lattice QCD computation
\cite{Isidori:2005tv}.

\begin{figure}[!t]
\begin{center}
\includegraphics[width=3.25cm]{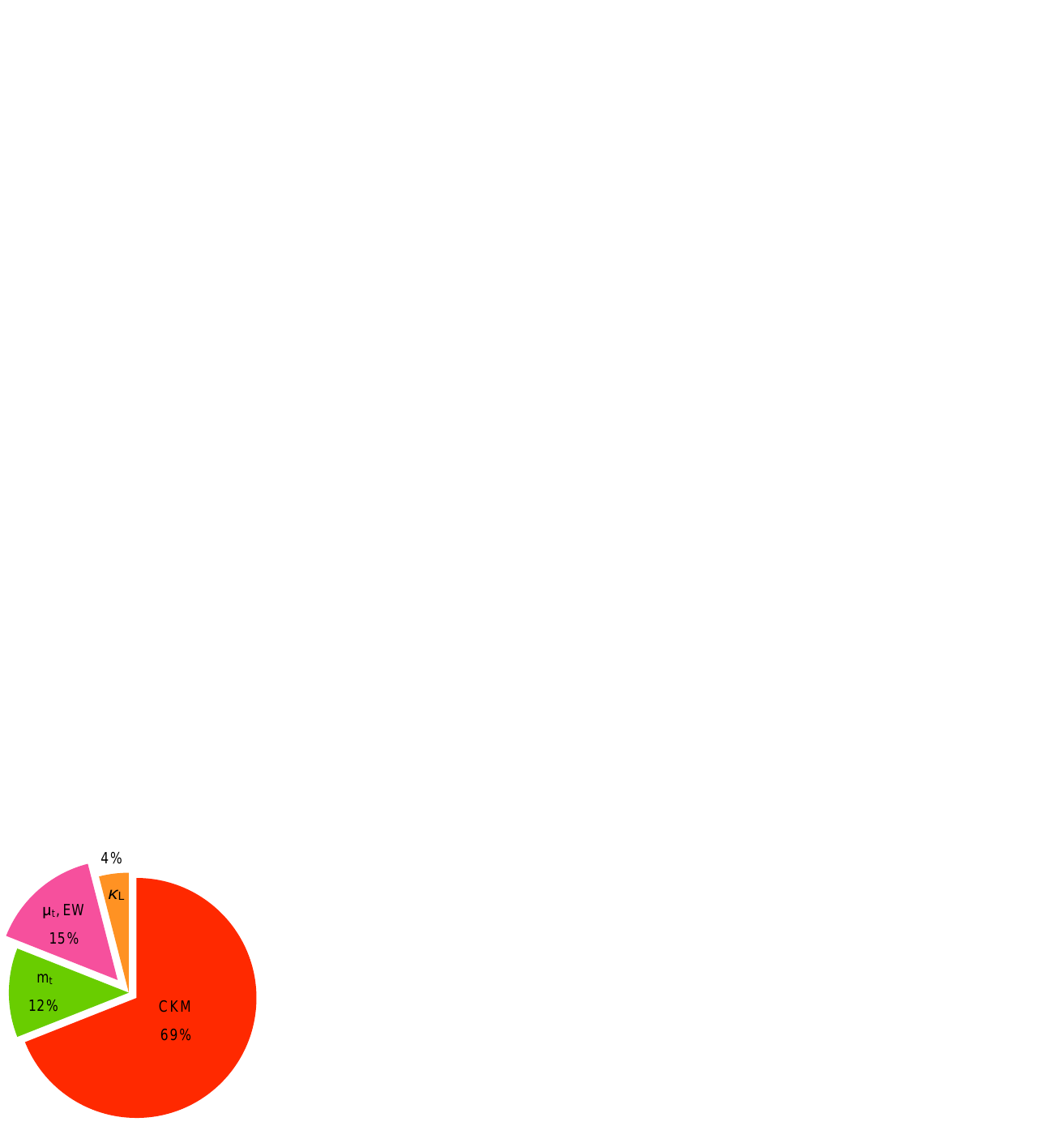}
\hspace{2cm}
\includegraphics[width=3.25cm]{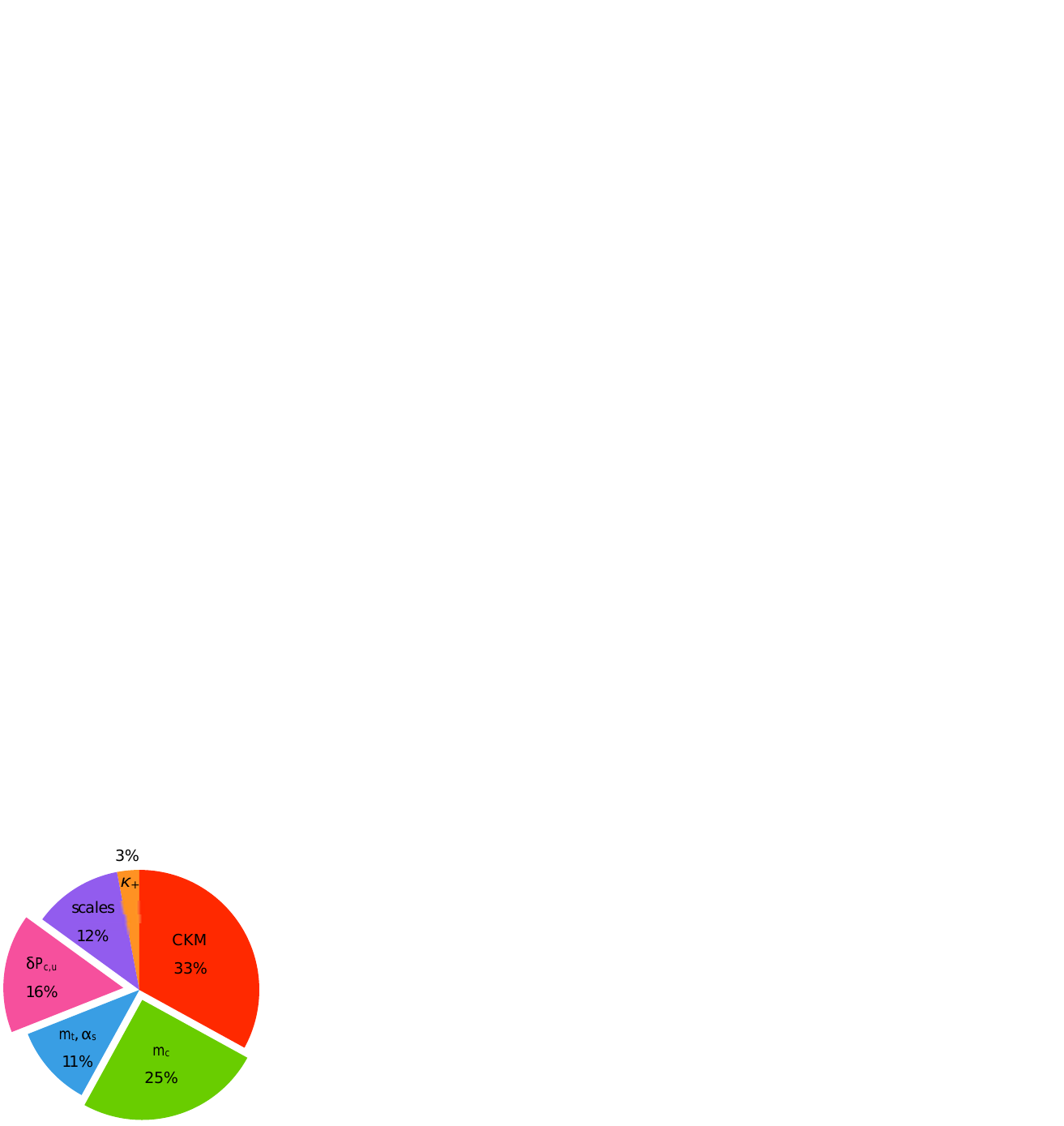}
\caption{Error budget of the SM prediction of $\BRKL$ (left) and
  $\BRKpg$ (right). See text for details.}
\label{fig:Kpies}
\end{center}
\end{figure}

Taking into account all the indirect constraints from the latest
unitarity triangle fit \cite{Charles:2004jd}, one finds, by adding
errors quadratically, the following SM predictions for the two
$\Knns$ rates
\begin{align} \label{eq:BRKSM}
\BRKL_{\rm SM} & = (2.54 \pm 0.35) \times 10^{-11} \, , \\
\BRKpg_{\rm SM} & = (7.96 \pm 0.86) \times 10^{-11} \, . 
\end{align}
The error budgets of the SM predictions of both $\Knns$ decays are
illustrated by the pie charts in \Fig{fig:Kpies}. As the breakdown of
the residual uncertainties shows, both decay modes are at present
subject mainly to parametric errors ($81 \%$, $69 \%$) stemming from
the CKM parameters, the quark masses $\mc$ and $\mt$, and $\as
(\MZ)$. The non-parametric errors in $\KLnn$ are dominated by the
uncertainty due to higher-order perturbative effects ($15 \%$), while
in the case of $\Kpnng$ the errors due to dimension-eight charm and LD
up quark effects ($16 \%$) and left over scale uncertainties ($12 \%$)
are similar in size. Given the expected improvement in the extraction
of the CKM elements through the $B$-factories, SM predictions for both
$\Knns$ rates with an accuracy significantly below $10 \%$ should be
possible before the end of this decade. Such precisions are unique in
the field of flavor-changing-neutral-current processes.

\section{\boldmath Recent theoretical progress in $\BXsga$ and $\BXsll$ \unboldmath} 

Considerable effort has gone into the calculation of fixed-order
logarithmic enhanced NNLO QCD corrections to $\BXsga$ \cite{NNLO,
  Q2ME, Misiak:2006ab, Misiak:2006zs}. A crucial part of the
NNLO calculation is the interpolation in the charm quark mass
performed in \cite{Misiak:2006ab}. The three-loop $\ord (\as^2)$
matrix elements of the current-current operators $Q_{1,2}$ contain the
charm quark, and the NNLO calculation of these matrix elements is
essential to reduce the overall theoretical uncertainty of the SM
calculation. In fact, the largest part of the theoretical uncertainty
in the NLO analysis of the BR is related to the definition of the mass
of the charm quark \cite{Gambino:2001ew} that enters the $\ord (\as)$
matrix elements $\langle s \gamma | Q_{1,2} | b \rangle$. The latter
matrix elements are non-vanishing at two loops only and the scale at
which $\mc$ should be normalized is therefore undetermined at
NLO. Since varying $\mc$ between $\mc (\mc) \sim 1.25 \, \GeV$ and
$\mc (\mb) \sim 0.85 \, \GeV$ leads to a shift in the NLO BR of more
than $10 \%$ this issue is not an academic one.
 
Finding the complete NNLO correction to $\langle s \gamma | Q_{1,2} |
b \rangle$ is a formidable task, since it involves the evaluation of
hundreds of three-loop on-shell vertex diagrams that are presently not
even known in the case $\mc = 0$. The approximation made in
\cite{Misiak:2006ab} is based on the observation that at the physical
point $\mc \sim 0.25 \, \mb$ the large $\mc \gg \mb$ asymptotic form
of the exact $\ord (\as)$ \cite{Buras:2002tp} and large-$\beta_0$
$\ord (\as^2 \beta_0)$ \cite{Q2ME} result matches the small
$\mc \ll \mb$ expansion rather well. This feature prompted the
analytic calculation of the leading term in the $\mc \gg \mb$
expansion of the three-loop diagrams, and to use the obtained
information to perform an interpolation to smaller values of $\mc$
assuming the $\ord (\as^2 \beta_0)$ part to be a good approximation of
the full $\ord(\as^2)$ result for vanishing charm quark mass. The
uncertainty related to this procedure has been assessed in
\cite{Misiak:2006ab} by employing three ans\"atze with different
boundary conditions at $\mc = 0$. A complete calculation of the $\ord
(\as^2)$ corrections to $\langle s \gamma | Q_{1,2} | b \rangle$ in
the latter limit or, if possible, for $\mc \sim 0.25 \, \mb$, would
resolve this ambiguity and should therefore be attempted.

Combining the aforementioned results it was possible to obtain the
first theoretical estimate of the total BR of $\BXsga$ at NNLO. For a
photon energy cut of $E_\gamma > E_{\rm cut}$ with $E_{\rm cut} = 1.6
\, \GeV$ in the $\bar{B}$-meson rest-frame the improved SM evaluation
is given by \cite{Misiak:2006ab, Misiak:2006zs}
\beq \label{eq:NNLO}
\BRga_{\rm SM} = (3.15 \pm 0.23) \times 10^{-4} \, , 
\eeq
where the uncertainties from hadronic power corrections ($5 \%$),
parametric dependences ($3 \%$), higher-order perturbative effects $(3
\%)$, and the interpolation in the charm quark mass ($3 \%$) have been
added in quadrature to obtain the total error.

The reduction of the renormalization scale dependence at NNLO is
clearly seen in the right panel of \Fig{fig:scales}. The most
pronounced effect occurs in the case of $\muc$ that was the main
source of uncertainty at NLO. The current uncertainty of $3 \%$ due to
higher-order effects is estimated from the variations of the NNLO
estimate under change of renormalization scales. The central value in
\Eq{eq:NNLO} corresponds to the choice $\mu_{{\scriptscriptstyle W},
  b, c} = (160, 2.5, 1.5) \, \GeV$. More details on the
phenomenological analysis including the list of input parameters can
be found in \cite{Misiak:2006ab}.

It is well-known that the operator product expansion (OPE) for
$\BXsga$ has certain limitations which stem from the fact that the
photon has a partonic substructure. In particular, the local expansion
does not apply to contributions from operators other than $Q_7$, in
which the photon couples to light quarks \cite{Kapustin:1995fk,
  lightquarks}. While the presence of non-local power corrections was
thus foreseen such terms have been studied until recently only in the
case of the $(Q_8, Q_8)$ interference \cite{Kapustin:1995fk}. In
\cite{Lee:2006wn} the analysis of non-perturbative effects that go
beyond the local OPE have been extended to the enhanced non-local
terms emerging from $(Q_7, Q_8)$ insertions. The found correction
scales like $\ord (\as \Lambda/\mb)$ and its effect on the BR was
estimated using the vacuum insertion approximation to be $-[0.3, 3.0]
\%$. A measurement of the flavor asymmetry between $\bar{B}^0 \to X_s
\gamma$ and $B^- \to X_s \gamma$ could help to sustain this numerical
estimate \cite{Lee:2006wn}. Potentially as important as the latter
corrections are those arising from the $(Q_{1,2}, Q_7)$
interference. Naive dimensional analysis suggests that some
non-perturbative corrections to them also scale like $\ord (\as
\Lambda/\mb)$. Since at the moment there is not even an estimate of
those corrections, a non-perturbative uncertainty of $5 \%$ has been
assigned to the result in \Eq{eq:NNLO}. This error is the dominant
theoretical uncertainty at present and thought to include all known
\cite{Lee:2006wn} and unknown $\ord (\as \Lambda/\mb)$
terms. Calculating the precise impact of the enhanced non-local power
corrections may remain notoriously difficult given the limited control
over non-perturbative effects on the light cone.

A further complication in the calculation of $\BXsga$ arises from the
fact that all measurements impose stringent cuts on the photon energy
to suppress the background from other $B$-meson decay
processes. Restricting $E_\gamma$ to be close to the physical endpoint
$E_{\rm max} = m_B/2$, leads to a breakdown of the local OPE, which
can be cured by resummation of an infinite set of leading-twist terms
into a non-perturbative shape function \cite{shapefunction}. A
detailed knowledge of the shape function and other subleading effects
is required to extrapolate the measurements to a region where the
conventional OPE can be trusted.

The transition from the shape function to the OPE region can be
described by a multi-scale OPE (MSOPE) \cite{Neubert:2004dd}. In
addition to the hard scale $\mu_h \sim \mb \sim 5 \, \GeV$, this
expansion involves a hard-collinear scale $\mu_{hc} \sim \sqrt{\mb
  \Delta} \sim 2.5 \, \GeV$ corresponding to the typical hadronic
invariant mass of the final state $X_s$, and a soft scale $\mu_s \sim
\Delta \sim 1.5 \, \GeV$ related to the width $\Delta/2 = \mb/2 -
E_{\rm cut}$ of the energy window in which the photon spectrum is
measured. In the MSOPE framework, the perturbative tail of the
spectrum receives calculable corrections at all three scales, and may
be subject to large perturbative corrections due to the presence of
terms proportional to $\as (\sqrt{\mb \Delta}) \sim 0.27$ and $\as
(\Delta) \sim 0.36$.

A systematic MSOPE analysis of the $(Q_7, Q_7)$ interference at NNLO
has been performed in \cite{Becher:2006pu}. Besides the hard matching
corrections, it involves the two-loop logarithmic and constant terms
of the jet \cite{Neubert:2004dd, jet} and soft function \cite{soft}.
The three-loop anomalous dimension of the shape function remains
unknown and is not included. The MSOPE result can be combined with the
fixed-order prediction by computing the fraction of events $1 - T$
that lies in the range $E_{\rm cut} = [1.0, 1.6] \, \GeV$. The
analysis \cite{Becher:2006pu} yields 
\beq \label{eq:T}
1 - T = 0.07{^{+0.03}_{-0.05}}_{\rm pert} \pm 0.02_{\rm hadr} \pm
0.02_{\rm pars} \, ,  
\eeq 
where the individual errors are perturbative, hadronic, and
parametric. The quoted value is almost twice as large as the NNLO
estimate $1 - T = 0.04 \pm 0.01_{\rm pert}$ obtained in fixed-order
perturbation theory \cite{Misiak:2006ab, Misiak:2006zs, Mikolaj} and
plagued by a significant additional theoretical error related to
low-scale perturbative corrections. These large residual scale
uncertainties indicate a slow convergence of the MSOPE series
expansion in the tail region of the photon energy spectrum. Given that
$\Delta$ is always larger than $1.4 \, \GeV$ and thus fully in the
perturbative regime this feature is unexpected.

Additional theoretical information on the shape of the photon energy
spectrum can be obtained from the universality of soft and collinear
gluon radiation. Such an approach can be used to predict large
logarithms of the form $\ln (E_{\rm max} - E_{\rm cut})$. These
computations have also achieved NNLO accuracy \cite{A&E} and
incorporate Sudakov and renormalon resummation via dressed gluon
exponentiation (DGE) \cite{A&E, Gardi:2006jc}. The present NNLO
estimate of $1 - T = 0.016 \pm 0.003_{\rm pert}$ \cite{A&E, Einan}
indicates a much thinner tail of the photon energy spectrum and a
considerable smaller perturbative uncertainty than reported in
\cite{Becher:2006pu}. The DGE analysis thus supports the view that the
integrated photon energy spectrum below $E_{\rm cut} = 1.6 \, \GeV$ is
well approximated by a fixed-order perturbative calculation,
complemented by local OPE power corrections. To understand how
precisely the tail of the photon energy spectrum can be calculated
requires nevertheless further theoretical investigations.

The study of $\btosll$ transitions can yield useful complementary
information, when confronted with the less rare $\btosg$ decays, in
testing the flavor sector of the SM. In particular, a precise
measurement of the inclusive $\BXsll$ decay distributions would be
welcome in view of NP searches, because they are amenable to clean
theoretical descriptions for dilepton invariant masses in the ranges
$1 \, \GeV^2 < q^2 < 6 \, \GeV^2$ \cite{Ghinculov:2003qd,
  Bobeth:2003at, Huber:2005ig} and $q^2 > 14 \, \GeV^2$
\cite{Ligeti:2007sn}. 

The SM calculations of the differential rate and forward-backward (FB)
asymmetry have both reached NNLO precision \cite{Ghinculov:2003qd,
  Bobeth:2003at, BXsllNNLO}. In the case of the the dilepton invariant
mass spectrum, integrated over the low-$q^2$ region, the most recent
SM prediction reads \cite{Huber:2005ig}
\beq \label{eq:BRll} 
\BRll_{\rm SM}^{1 \, \GeV^2 < q^2 < 6 \, \GeV^2}
= (1.59 \pm 0.11) \times 10^{-6} \, . 
\eeq
The position of the zero of the FB asymmetry is known to be especially
sensitive to NP effects. In the SM one finds \cite{Bobeth:2003at}
\beq \label{eq:AFBq02}
q_{0, {\rm SM}}^2 = (3.76 \pm 0.33) \, \GeV^2 \, . 
\eeq
The total errors in \Eqsand{eq:BRll}{eq:AFBq02} have been obtained by
adding the individual parametric and perturbative uncertainties in
quadrature. Besides the differential rate and FB asymmetry, an angular
decomposition of $\BXsll$ provides a third observable that is
sensitive to a different combination of Wilson coefficients. This
recent observation \cite{Lee:2006gs} might allow to extract SD
information from limited data on $\BXsll$ in the low-$q^2$ region more
efficiently.

Like in the case of $\BXsga$, experimental cuts complicate the
theoretical description of $\BXsll$ as they make the measured decay
distributions sensitive to the non-perturbative shape function
\cite{MXcut}. In particular, putting an upper cut $M_{X_s}^{\rm cut} =
[1.8, 2.0] \, \GeV$ on the hadronic invariant mass of $X_s$
\cite{bxsll}, in order to suppress the background from $\bar{B} \to
X_c \ell \bar{\nu} \to X_s \ell^+ \ell^- \nu \bar{\nu}$ at small
$q^2$, causes a reduction of the rate by $(10 - 30) \%$ \cite{MXcut}.
Although the reduction can be accurately calculated using the
universal $\BXsga$ shape function \cite{MXcut}, subleading shape
functions may introduce an additional error of $5 \%$ in
\Eq{eq:BRll}. Similarly, no additional uncertainty for unknown
subleading non-perturbative corrections has been included in
\Eqsand{eq:BRll}{eq:AFBq02}. Most importantly, uncalculated $\ord (\as
\Lambda/\mb)$ non-perturbative corrections may imply an additional
uncertainty of $5 \%$ in the above formulas. This issue deserves an
dedicated study.

\section{\boldmath CMFV: combining $\Kpnng$, $\BXsga$, $\BXsll$, and
  $\Ztobb$}

A way to study possible NP effects in flavor physics consists in
constraining the Wilson coefficients of the operators in the
low-energy effective theory. This model-independent approach to MFV
has been applied combining various $K$- and $B$-meson decay modes both
including \cite{MFV} and neglecting \cite{CMFV, Bobeth:2005ck,
  Haisch:2007ia} operators that do not contribute in the SM, that is,
so-called constrained MFV (CMFV) \cite{Blanke:2006ig} scenarios.

\begin{figure}[!t]
\vspace{-0.5cm}
\begin{center}
\includegraphics[width=7cm,height=7cm]{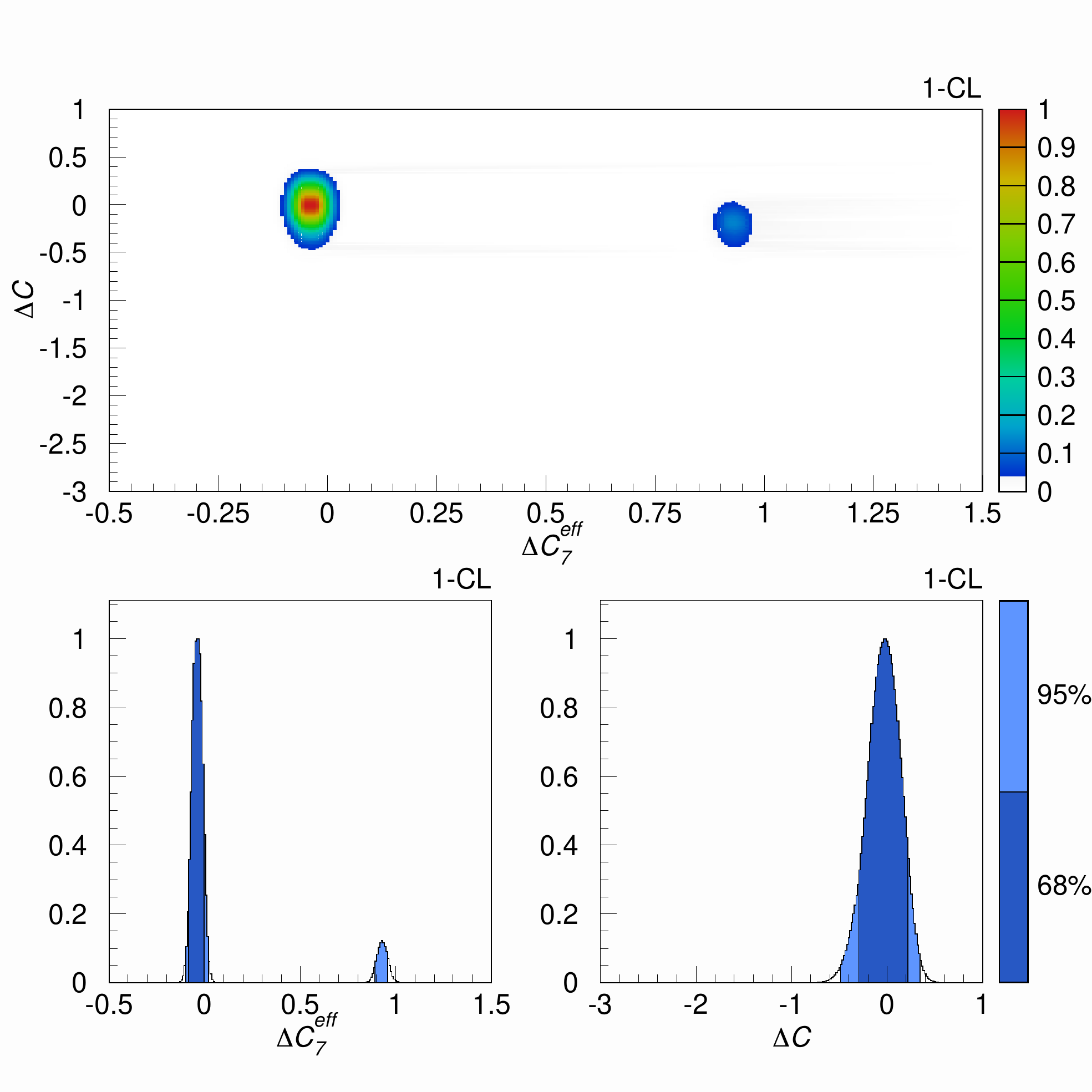}
\hspace{5mm}
\includegraphics[width=7cm,height=7cm]{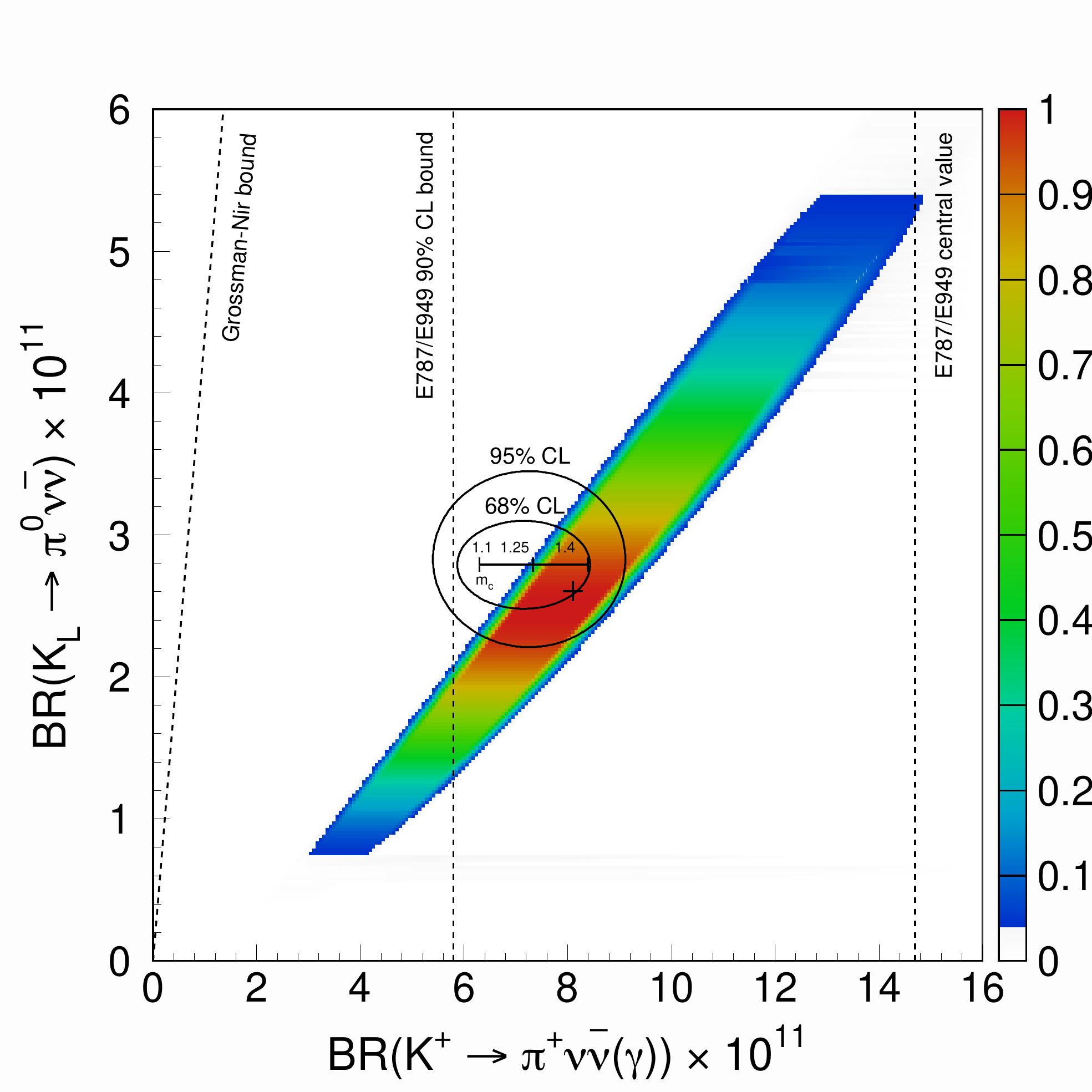}
\vspace{-2mm}
\caption{Left: Constraints on $\Delta C_7^{\rm eff}$ and $\Delta C$
  within CMFV that follow from a combination of the measurements of
  $\BXsga$, $\BXsll$, and $\Ztobb$. The colors encode the frequentist
  $1 - {\rm CL}$ level and the corresponding $68 \% \, {\rm CL}$ and
  $95 \% \, {\rm CL}$ regions as indicated by the bars on the right
  side of the panel. Right: Frequentist $1 - {\rm CL}$ level of
  $\BRKpg$ {\it vs}. $\BRKL$ in CMFV. The two black ellipses indicate
  the $68 \% \, {\rm CL}$ and $95 \% \, {\rm CL}$ regions in the SM
  while the best fit value within CMFV is specified by the black
  cross. The central value of $\BRKpg_{\rm SM}$ as a function of the
  charm quark $\MSbar$ mass in $\GeV$ and the lower experimental $90
  \% \, {\rm CL}$ bound and central value are also shown. See text for
  details.}
\label{fig:C7vsC&KpvsKL}
\end{center}
\end{figure}

The main goal of the recent CMFV study \cite{Haisch:2007ia} was an
improved determination of the range allowed for the NP contribution
$\Delta C = C - C_{\rm SM}$ to the universal $Z$-penguin by a thorough
global fit to the $\Ztobb$ pseudo observables (POs) $\Rb$, $\Ab$, and
$\AFB$ \cite{ewpm} and the measured $\BXsga$ \cite{bsgamma} and
$\BXsll$ \cite{bxsll} rates. The derived model-independent bounds
\beq \label{eq:dcsb0} 
\Delta C = -0.026 \pm 0.264 \;\; (68 \% \, {\rm CL}) \, , \hspace{5mm}
\Delta C = [-0.483, 0.368] \;\; (95 \% \, {\rm CL}) \, ,
\eeq 
imply that large negative contributions that would reverse the sign of
the SM $Z$-penguin amplitude are highly disfavored in CMFV scenarios
due to the strong constraint from $\Rb$
\cite{Haisch:2007ia}. Interestingly, such a conclusion cannot be drawn
by considering only flavor constraints \cite{Bobeth:2005ck}, since a
combination of $\BRga$, $\BRll$, and $\BRKpg$ does not allow to
distinguish the SM solution $\Delta C = 0$ from the wrong-sign case
$\Delta C \sim -2$ at present. The constraint on $\Delta C$ within
CMFV following from the simultaneous use of $\Rb$, $\Ab$, $\AFB$,
$\BRga$, and $\BRll$ can be seen in the left panel of
\Fig{fig:C7vsC&KpvsKL}.

One can also infer from this figure that two regions, resembling the
two possible signs of the amplitude $A (\btosg) \propto C_7^{\rm eff}
(\mb)$, satisfy all existing experimental bounds. The best fit value
for the NP contribution $\Delta C_7^{\rm eff} = C_7^{\rm eff} (\mb) -
C_{7 {\rm SM}}^{\rm eff} (\mb)$ is very close to the SM point residing
in the origin, while the wrong-sign solution located on the right is
highly disfavored, as it corresponds to a $\BRll$ value considerably
higher than the measurements \cite{Gambino:2004mv}. The corresponding
limits are \cite{Haisch:2007ia}
\beq \label{eq:dc7effsb0}
\Delta C_7^{\rm eff} = -0.039 \pm 0.043 \;\; (68 \% \, {\rm CL}) \, ,
\hspace{2.5mm} \Delta C_7^{\rm eff} = [-0.104, 0.026] \, \cup \,
[0.890, 0.968] \;\; (95 \% \, {\rm CL}) \, . 
\eeq 
Similar bounds have been presented previously in
\cite{Bobeth:2005ck}. Since $\BRga_{\rm SM}$ as given in \Eq{eq:NNLO}
is lower than the experimental world average $\BRga = (3.55 \pm 0.26)
\times 10^{-4}$ \cite{Barbiero:2007cr} by $1.2 \mysigma$, extensions
of the SM that predict a suppression of the $\btosg$ amplitude are
strongly constrained. In particular, even the SM point $\Delta
C_7^{\rm eff} = 0$ is almost disfavored at $68 \% \, {\rm CL}$ by the
global fit.

The stringent bound on the NP contribution $\Delta C$ given in
\Eq{eq:dcsb0} translates into tight two-sided limits for the BRs of
all $Z$-penguin dominated $K$- and $B$-decays \cite{Haisch:2007ia}. In
the case of the $\Knns$ modes the allowed ranges for the CMFV BRs read
\cite{Haisch:2007ia}
\beq \label{eq:CMFVbounds}
\begin{split}
  \BRKL_{\rm CMFV} & = [1.55, 4.38] \times 10^{-11}~~~(95 \% \, {\rm
    CL}) \, , \\ \BRKpg_{\rm CMFV} & = [4.29, 10.72] \times
  10^{-11}~~~(95 \% \, {\rm CL}) \, ,
\end{split}
\eeq
when NP contributions to EW boxes are neglected. These bounds imply
that in CMFV models the $\KLnn$ and $\Kpnng$ rates can differ by at
most $^{+18 \%}_{-21 \%}$ and $^{+27 \%}_{-30 \%}$ from their SM
expectations. In particular, the Grossman-Nir bound $\BRKL \lesssim
4.4 \, \BRKpg$ \cite{Grossman:1997sk} following from isospin
symmetry, cannot be saturated in CMFV models. This is illustrated by
the plot in the right panel of \Fig{fig:C7vsC&KpvsKL} which shows the
frequentist $1 - {\rm CL}$ level of the prediction $\BRKpg_{\rm
  CMFV}$ {\it vs}. $\BRKL_{\rm CMFV}$.

A strong violation of the CMFV bounds in \Eq{eq:CMFVbounds} by future
precision measurements of $\BRKL$ and/or $\BRKpg$ will imply a failure
of the CMFV assumption, signaling either the presence of new effective
operators and/or new flavor and $\CP$ violation. A way to evade the
above limits is the presence of sizable EW box contributions. While
these possibility cannot be fully excluded, general arguments and
explicit calculations indicate that it is difficult to realize in the
CMFV framework.

\end{document}